\documentclass[a4paper,10pt]{scrartcl}

\usepackage{latexsym}
\usepackage{amsmath}
\usepackage{amsfonts}
\usepackage{amssymb}
\usepackage{graphicx}
\usepackage[english]{babel}
\usepackage{wasysym}
\usepackage{units}
\newcommand{\p}{\partial}
\newcommand{\reff}[1]{(\ref{#1})}
\newcommand{\vs}[1]{\vspace{#1mm}}

\newcommand{\vsO}{\vspace{.1cm}\hfill\\}
\newcommand{\vsT}{\vspace{.2cm}\hfill\\}

\title{\Large THE JEANS MECHANISM AND\\ BULK-VISCOSITY EFFECTS}

\author{Nakia Carlevaro$^{\;a,\;b}$ and Giovanni Montani$^{\;b,\;c,\;d,\;e}$\vsT
\emph{\footnotesize $^a$Department of Physics, Polo Scientifico -- Universit\`a degli Studi di Firenze,}\vs{-2.5}\\
\emph{\footnotesize INFN -- Section of Florence, Via G. Sansone, 1 (50019), Sesto Fiorentino (FI), Italy}\\
\emph{\footnotesize $^b$ICRA -- International Center for Relativistic Astrophysics,}\vs{-2.5}\\
\emph{\footnotesize c/o Dep. of Physics - ``Sapienza'' Universit\`a di Roma}\\
\emph{\footnotesize $^c$ Department of Physics - ``Sapienza'' Universit\`a di Roma, Piazza A. Moro, 5 (00185), Rome, Italy}\\
\emph{\footnotesize $^d$ENEA -- C.R. Frascati (Department F.P.N.), Via Enrico Fermi, 45 (00044), Frascati (Rome), Italy}\\
\emph{\footnotesize $^{e}$ ICRANet -- C. C. Pescara, Piazzale della Repubblica, 10 (65100), Pescara, Italy}\vsO
{\footnotesize\ttfamily nakia.carlevaro@icra.it\quad montani@icra.it}
}
\date{}
\begin{document}
\maketitle

%
\hrule
\begin{abstract} \textbf{Abstract:} In this paper we study the gravitational instability in presence of viscosity. In particular, the standard Jeans Mechanism is analyzed taking into account bulk-viscosity effects in the first-order Newtonian dynamics. We show how the perturbation evolution is damped by dissipative processes and the top-down fragmentation mechanism is suppressed for strong viscous effects. The critical value of the Jeans Mass remains unchanged also in presence of bulk viscosity.
\end{abstract}
\hrule

\vspace{1cm}
\section{Introduction}
The crucial dichotomy between the isotropy of region at red-shift $z_{rs}\sim10^{3}$ and the extreme irregularity of the recent Universe, $z_{rs}\ll1$, is at the ground of the interest in the study of the gravitational instability for the structure formation. The fundamental result of the cosmological density-perturbation analysis is the so-called \emph{Jeans Mass}, which is the threshold value for the fluctuation masses to condense, generating a real structure. If masses greater than the Jeans one are addressed, density perturbations begin to diverge as function of time giving rise to the gravitational collapse \cite{jeans02,jeans}. In this work, we consider dissipative effects into the fluid dynamics, in the linear Newtonian regime. The starting point is the fluid motion equations on which one can develop a first-order perturbative theory by adding small fluctuations to the unperturbed background. In particular, we introduce in the first-order analysis the so-called \emph{bulk viscosity}, described as a function of the Universe energy density $\rho$ via a power-law: $\zeta=z\,\rho^{\,s}$ where $z,s=const.$ We show how the perturbations are damped by viscosity and the top-down fragmentation mechanism is suppressed. The critical value of the Jeans Mass remains unchanged also in presence of bulk viscosity. In the the standard \emph{Jeans Mechanism} \cite{jeans02,weinberg,w71}, the unperturbed background is assumed to be characterized by a static and uniform solution of the fluid parameters.
\newcommand{\vo}{\textbf{v}_{\scriptscriptstyle 0}}
\newcommand{\ro}{\rho_{\scriptscriptstyle 0}}
\newcommand{\rb}{\bar{\rho}}
\newcommand{\gb}{\bar{\gamma}}
\newcommand{\wb}{\bar{\omega}}
\newcommand{\po}{p_{\scriptscriptstyle 0}}
\newcommand{\phio}{\phi_{\scriptscriptstyle 0}}
\newcommand{\zetao}{\zeta_{\scriptscriptstyle 0}}
\newcommand{\zetaob}{\bar{\zeta}_{\scriptscriptstyle 0}}
\newcommand{\w}{\omega}
\newcommand{\ao}{a_{\scriptscriptstyle 0}}
\newcommand{\vu}{\textbf{v}_{\scriptscriptstyle 1}}
\newcommand{\ru}{\rho_{\scriptscriptstyle 1}}
\newcommand{\pu}{p_{\scriptscriptstyle 1}}
\newcommand{\phiu}{\phi_{\scriptscriptstyle 1}}
\newcommand{\zetau}{\zeta_{\scriptscriptstyle 1}}
\newcommand{\vup}{\dot{\textbf{v}}_{\scriptscriptstyle 1}}
\newcommand{\rup}{\dot{\rho}_{\scriptscriptstyle 1}}

\section{Motion Equations of Viscous Fluids}
In order to describe the Newtonian evolution of a fluid, we here want to introduce the Eulerian equations governing the fluid parameters: the density $\rho$, the local 3-velocity $\textbf{v}$ (of components $v_\alpha$) and the pressure $p$, in presence of a gravitational potential $\phi$ (indices $\alpha$ and $\beta$ run from 1 to 3). Since we are interested to treat isotropic and homogeneous perturbative cosmological models, we can safely neglect the so-called \emph{first viscosity} (\emph{shear viscosity}) in the unperturbed dynamics. In fact, in such models there is no displacement of matter layers with respect to each other and this kind of viscosity represents the energy dissipation due to this particular effect. Indeed, in presence of small inhomogeneities, such affect should be taken into account, in principle. However, in this work, we are aimed at studying the behavior of scalar density perturbations. In this respect, volume changes of a given mass scale are essentially involved and, therefore, we concentrate our attention to the effects of the so-called \emph{second viscosity} or \emph{bulk viscosity}, only. In fact, such kind of viscosity is responsible for the non-equilibrium dynamics of matter compression and rarefaction and we expect that such a phenomenon is more relevant than friction among different layers. According to literature developments \cite{bk76,bk77,bnk79,lk63,b87,b-arxiv,paddy}, we now assume the bulk-viscosity coefficient $\zeta$ as a function of the energy density $\rho(t)$ expressed via a power-law of the form
\begin{equation}\label{bulk-power-law}
\zeta=z\,\rho^{\,s}\;,\qquad\qquad z,s=const.\;,
\end{equation}
here $z$ is a constant parameter which defines the intensity of viscous effects \cite{b88}.

Viscous fluids are governed, in Newtonian regime, by the following set of equations \cite{landau-fluid}: 
\begin{subequations}\label{newtonian-motion-eq}
\begin{align}
\label{continuity-eq}
\p_t{\rho}+\nabla\cdot(\rho\textbf{v})&=0\;,\\
\label{navier-stokes-eq}
\rho\,\p_t{\textbf{v}}+ \rho\,(\textbf{v}\cdot\nabla)\textbf{v} + \nabla p\,-
\zeta\,\nabla(\nabla \cdot \textbf{v})+\rho\,\nabla\phi&=0\;,\\
\label{poisson-eq}
\nabla^2 \phi-4\pi G \rho &=0\;,
\end{align}
\end{subequations}
which are the \emph{Continuity Eq.} (energy conservation), the \emph{Navier-Stokes Eq.} (momentum conservation) and the \emph{Poisson Eq.} for the gravitational field $\phi$, respectively. The pressure and density are linked by the \emph{Eq. of State} (EoS), \emph{i.e.}, $p=p\,(\rho)$: in this picture, the sound speed is defined by the relation $v_s^2=\nicefrac{\delta p}{\delta\rho}$.

\section{Analysis of the dissipative Jeans Mechanism}
The mass agglomerates are due to the gravitational instability: if density perturbations are generated in a certain volume, the gravitational forces act contracting this volume, allowing a gravitational collapse. The only forces which contrast such gravitational contraction are the pressure ones. The Jeans Mechanism \cite{jeans02} analyzes what are the conditions for which density perturbations become unstable to the gravitational collapse. Such a model is based on a Newtonian approach and the effects of the expanding Universe are neglected. The fundamental hypothesis is the static and uniform solution for the zeroth-order dynamics
\begin{equation}\label{static-swindle-solution}
\vo=0\;,\qquad\ro=const.\;,\qquad \po=const.\;,\qquad\phio=const.
\end{equation}
Of course, this assumption contradicts the gravitational equation, but we follow the original Jeans analysis imposing the so-called ``Jeans swindle''\cite{jeans,weinberg}. We underline that our study will focus on Universe stages when the mean density is very small: in particular the recombination era, after decoupling. This way, the effects of bulk viscosity on the unperturbed dynamics can be consistently neglected in view of its phenomenological behavior \reff{bulk-power-law}.

\subsection{Review of the Jeans Mechanism} In the original Jeans Model, a perfect fluid background is assumed. After setting $\zeta=0$ in eqs. \reff{newtonian-motion-eq}, let us add small fluctuations to the unperturbed solution: $\rho=\ro+\delta\rho,\;p=\po+\delta p$, $\phi=\phio+\delta\phi$, and $\textbf{v}=\vo+\delta\textbf{v}$. Substituting such expressions, one differential equation for the density perturbations can be derived:
\begin{equation}\label{eqdiffjeans}
\p_t^2{\delta\rho}-v_s^2\;\nabla^2\delta\rho=4\pi G\,\ro\,\delta\rho\;.
\end{equation}
To study the properties of $\delta\rho$, we now consider a plane-wave solutions of the form
\begin{equation}\label{plane-wave}	
\delta\rho\,(\textbf{r},t) = A\;e^{i\omega t - i\textbf{k}\cdot\textbf{r}}\;,
\end{equation}
where $\omega$ and $\textbf{k}$ ($k=|\textbf{k}|$) are the angular frequency and the wave number, respectively. This way, one can obtain the following dispersion relation
\begin{equation}\label{dispersion-jeans}
\omega^2=v_s^2k^2-4\pi G\,{\ro}\;.
\end{equation}

In this scheme, two different regimes are present: if $\omega^{2}>0$ a pure time oscillatory behavior for density perturbations is obtained. While if $\omega^{2}<0$, the fluctuations exponentially grow in time, in the $t\to\infty$ asymptotic limit (\emph{i.e.}, we choose the negative imaginary part of the angular-frequency solution) and the gravitational collapse is addressed since also the density contrast $\delta=\nicefrac{\delta\rho}{\ro}$ diverges. The condition $\omega^2=0$ defines the so-called \emph{Jeans Scale} $K_J$ and the Jeans Mass $M_J$ (which is the total mass in a sphere of radius $R=\pi/K_J$). Such threshold quantities read
\begin{equation}\label{jeans-mass}
K_J=\rho_{\scriptscriptstyle 0}\sqrt{\frac{4\pi G
\rho_{\scriptscriptstyle 0}}{v_s^2}}\;,\qquad
M_J=\frac{4\pi}{3}\left(\frac{\pi}{K_J}\right)^3
\rho_{\scriptscriptstyle 0}=
\frac{\pi^{5/2}\,v_s^3}{6G^{3/2}{\rho_{\scriptscriptstyle 0}}^{1/2}}\;.
\end{equation}

\subsection{Viscous effects} Let us now analyze how bulk viscosity can affect the gravitational collapse dynamics and we recall that we are able to neglect such kind of viscosity in the unperturbed dynamics, which results to be described by the static and uniform solution \reff{static-swindle-solution}. We start by adding the usual small fluctuations to such a solution, \emph{i.e.}, $\delta\rho,\;\delta p,\;\delta\phi,\;\delta\textbf{v}$. In treating bulk-viscosity perturbations, we use a standard Taylor expansion $\zeta=\zetao+\delta\zeta$ where $\zetao=\zeta(\ro)$. Substituting all fluctuations in the system \reff{newtonian-motion-eq}, one can obtain an unique equation for density perturbations, describing the dynamics of the gravitational collapse:
\begin{equation}\label{eqfond3.1}
\ro\;\p_t^2{\delta\rho}-\ro\;v_s^2\nabla^2\,\delta\rho-
\zetao\,\nabla^2\,\p_t{\delta\rho}=4\pi G\ro^2\;\delta\rho\;.
\end{equation}
Plane-waves solutions \reff{plane-wave} can be addressed, obtaining a generalized dispersion relation
\begin{equation}
\ro\,\omega^2-\,i\,\zetao\,k^2\;\omega+\ro(4\pi G\ro-v_s^2 k^2)=0\;.
\end{equation}

As in the standard Jean Model, the nature of the angular frequency is responsible of two different regimes. The dispersion relation has the solution 
\begin{equation}
\omega=i(\zetao k^2)/(2\,\ro)\pm
\sqrt{\bar{\omega}}\;,\qquad\;\;
\bar{\omega}=-(k^4\zetao^2)/(4\ro^2)+v_s^2k^2-4\pi G\ro\;,
\end{equation}
thus we obtain the time exponential regime for $\bar{\omega}\leqslant0$ and a damped oscillatory regime for $\bar{\omega}>0$. The equation $\bar{\omega}=0$ admits the solutions $K_1$ and $K_2$ which read 
\begin{equation}
K_{1,2}=(\sqrt{2}\,\ro v_s)/(\zetao)
\,\big(1\mp\sqrt{1-(K_{J}\zetao)/(\ro v_s)^{2}}\;\;\big)^{1/2}\;,
\quad
K_{1},K_{2}>0,\quad K_{1}<K_{2}\;.
\end{equation}
The existence of such solutions gives rise to a first constraint for the viscosity coefficient: $\zetao\leqslant\zeta_c=\ro v_s/K_J$. An estimation in the recombination era\footnote{\scriptsize The parameters are set as follows: the usual barotropic relation $p=c^2\ro^\gamma/\tilde{\rho}^{\gamma-1}$ is assumed and the constant $\tilde{\rho}$ can be derived from the expression expression $M_J$ \reff{jeans-mass}. Universe is dominated by matter and we can impose the values: $M_J\sim 10^6 M_\odot$, $\gamma=5/3$, $\rho_{c}=1.879 \,h^2 \cdot 10^{-29} \;g\,cm^{-3}$, $h=0.7$, $z=10^{3}$ and $\ro=\rho_{c}\,\,z^3 = 0.92\cdot10^{-20} \;g\,cm^{-3}$\;. Using these quantities one finds $\tilde{\rho}= 9.034 \cdot 10^{-7}	\;g\,cm^{-3}$, $v_s=8.39\cdot 10^5\;cm\,s^{-1}$ and the threshold value $\zeta_c$.}, after decoupling, yields to the value $\zeta_c= 7.38 \cdot 10^{4}	\;g\,cm^{-1}\,s^{-1}$ and confronting this threshold with usual viscosity (\emph{e.g.}, $\zetao^{Hydr.}=8.4\cdot10^{-7}g\,cm^{-1}\,s^{-1}$), we can conclude that the range $\zetao\leqslant\zeta_c$ is the only of physical interest. Finally we obtain: $\bar{\omega}\leqslant0$ for $k\leqslant K_1$, $K_2\leqslant k$ and $\bar{\omega}>0$ for $K_1<k<K_2$.

Let us now analyze the density-perturbation exponential solutions for $\bar{\omega} \leqslant 0$:
\begin{equation}
\delta\rho\sim e^{\textrm{w}\,t}\;,\qquad\quad
\textrm{w}=-(\zetao k^2)/(2\ro)\mp\sqrt{-\bar{\omega}}\;.
\end{equation}
To obtain the structure formation, the amplitude of such stationary waves must grow for increasing time. The exponential collapse for $t\to\infty$ is addressed, choosing the $(+)$ sign solution, only if $\textrm{w}>0$, \emph{i.e.}, $k<K_J$ with $K_J<K_1<K_2$. As a result, we show how the structure formation occurs only if $M>M_J$, as in the standard Jeans Model. The viscous effects do not alter the threshold value of the Jeans Mass, but they change the perturbation evolution and the pure oscillatory behavior is lost in presence of dissipative effects. In particular, we get two distinct decreasing regimes: for $K_1<k<K_2$ (\emph{i.e.}, $\wb>0$), we obtain a damped oscillatory evolution of perturbations:
$\delta\rho\sim \textrm{exp}[-t(\zetao k^2)/(2\ro)]\;\cos{(\sqrt{\bar{\omega}}\;t)}\;,$ while, for $K_J<k<K_1$ and $K_2<k$, density perturbations exponentially decrease as $\delta\rho\sim e^{\textrm{w}\,t}$, with $\textrm{w}<0$, in the limit $t\to\infty$.

\section{Implication for the top-down mechanism}
As shown above, since the pure oscillatory regime does not occurs, we deal with a decreasing exponential or a damped oscillatory evolution of perturbations. This allows to perform a qualitative analysis of the top-down fragmentation scheme \cite{kolb-turner}, \emph{i.e.}, the comparison between the evolution of two structures: one collapsing agglomerate with $M_G\gg M_J$ and an internal non-collapsing sub-structure with $M_S<M_J$. If this picture is addressed, the sub-structure mass must be compared with a decreasing Jeans Mass since the latter is inversely proportional to the collapsing-agglomerate background mass. This way, as soon as such a Jeans Mass reaches the sub-structure one, the latter begins to condense implying the fragmentation. In the standard Jeans Model, this mechanism is always allowed since the amplitude for perturbations characterized by $M_S<M_J$ remains constant in time. On the other hand, the presence of decreasing fluctuations in the viscous model, requires a discussion on the effective damping and an the efficacy of the top-down mechanism. Of course, such an analysis contrasts the hypothesis of a constant background density, but it can be useful to estimate the strength of the dissipative effects. We now study two cases for different values of the bulk-viscosity coefficient: $\zetao\ll1$ and $\zetao>1$. In this analysis, a perturbative validity-limit has to be set: we suppose $\nicefrac{\delta\rho}{\ro}\sim0.01$ as the limit of the model and we use the recombination era parameters (see footnote $^{1}$), in particular the initial time of the collapse is define as the beginning of the matter-dominated Universe, \emph{i.e.}, $t_{\scriptscriptstyle 0}=t_{MD}=1.39\cdot10^{13}\,s$. 

In correspondence of a very small viscosity coefficient (Fig.1),
\begin{figure}[!ht]
\begin{minipage}{6.7cm}
\centering
\includegraphics[width=6.7cm]{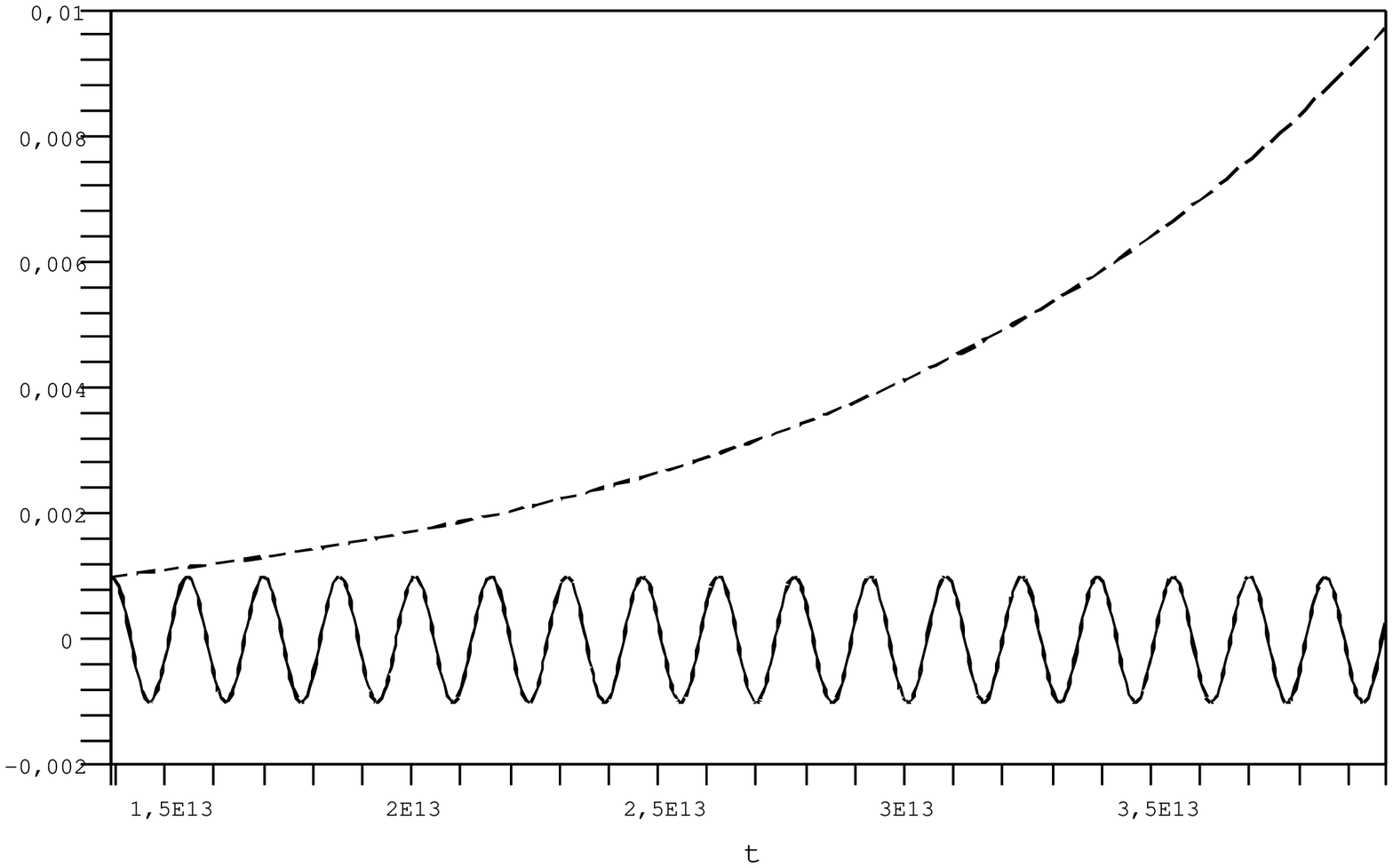}
\caption{\scriptsize{Case $\zetao=14\,g\,cm^{-1}\,s^{-1}$.  Galaxy density contrast: $\delta_G$ - $M_G=10^{12}\,M_\odot$ - (dashed line). Sub-structure density contrast $\delta_S$ - $M_S=M_\odot$ - (normal line).}}
\end{minipage}
\ \hspace{1mm} \hspace{0.5mm} \
\begin{minipage}{6.7cm}
\centering
\includegraphics[width=6.7cm]{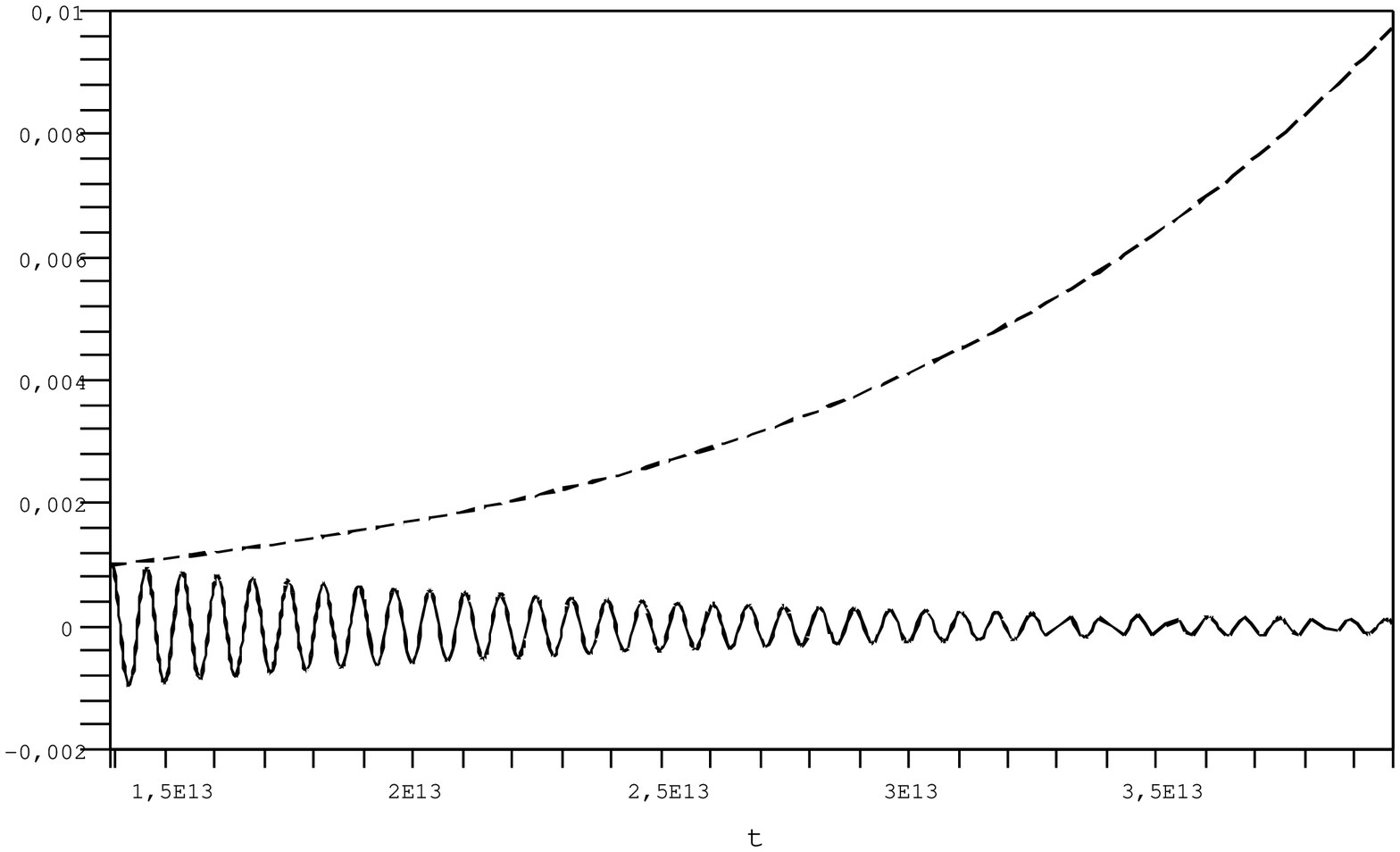}
\caption{\scriptsize{Case $\zetao=10^{-5}\,g\,cm^{-1}\,s^{-1}$.  Galaxy density contrast: $\delta_G$ - $M_G=10^{12}\,M_\odot$ - (dashed line). Sub-structure density contrast $\delta_S$ - $M_S=10\,M_\odot$ - (normal line).}}
\end{minipage}
\end{figure}
we consider the Jeans Mass $M_J=10^6\,M_\odot$ and a sub-structure wave-number $K_S$ in the region $K_1<K_S<K_2$, \emph{i.e.}, density perturbations evolve like damped oscillations. Fluctuations have to be imposed small at the initial time $t_{\scriptscriptstyle 0}$, this way, we consider density contrasts ($\delta_G$ for the galaxy and $\delta_S$ for the sub-structure) of $\mathcal{O}(10^{-3})$. In this scheme, the galaxy starts to collapse and the validity limit is reached at $t^*=6.25\cdot10^{13}$. As a result, in Fig.1 we can show how the sub-structure survives during the background collapse until the threshold time value $t^*$. Thus, we can conclude that, if the viscous damping is sufficiently small, the galaxy formation occurs.

Let us now discuss the case $\zetao>1$ (Fig.2) by changing the sub-structure mass, which is now $M_S=M_\odot$. Here, the damping effects is stronger. In fact, when the galaxy density contrast reaches the threshold value $\delta_G=0.1$, we obtain $\delta_S=10^{-5}$. The top-down mechanism for structure formation results to be unfavored by the presence of strong viscous effects: the damping becomes very strong and the sub-structure vanishes during the agglomerate evolution.

\section{Concluding remarks}
The effects induced by the presence of bulk viscosity on the Jeans Mechanism have been analyzed in a perturbative scheme. Viscosity has been introduced in the first-order dynamics via a power-low function of the energy density and the effects produced on the unperturbed dynamics have been consistently neglected in view of the phenomenological nature of such kind of viscosity.

The main result has been to show how bulk viscosity damps the density contrast evolution maintaining unchanged the threshold value of the Jeans Mass. Such an effect suppresses the sub-structure formation in the top-down fragmentation mechanism. In particular, a new decreasing regime for perturbations has been found. The presence of such a behavior allowed the study of the top-down scheme for small and strong viscous effects. In the first case, the density-perturbation amplitude of a sub-structure remains substantially constant during the main structure collapse. On the other hand, if viscous effects are sufficiently strong, the sub-structure vanish in the linear perturbative regime, unfavoring the fragmentation.


\begin{thebibliography}{99} 

\bibitem{jeans02}
J.H. Jeans, 
{\it Phil. Trans. Roy. Soc. London A} {\bf 199}, 1 (1902).

\bibitem{jeans}
J.H. Jeans, 
{\it Astronomy and Cosmogony} (Cambridge Uni. Press, Cambridge (UK) 1928).

\bibitem{weinberg}
S. Weinberg,
{\it Gravitation and Cosmology: Principles and Applications of the General Theory of Relativity} (John Wiley \& Sons, New York (USA), 1972).

\bibitem{w71}
S. Weinberg, 
{\it Atrophys. J.} {\bf 168}, 175 (1971). 

\bibitem{bk76}
V.A. Belinskii and I.M. Khalatnikov, 
{\it Sov. Phys. JETP} {\bf 42}(2), 205 (1976).

\bibitem{bk77}
V.A. Belinskii and I.M. Khalatnikov, 
{\it Sov. Phys. JETP} {\bf 45}(1), 1 (1977).

\bibitem{bnk79}
V.A. Belinskii, E.S. Nikomarov, I.M. Khalatnikov, 
{\it Sov. Phys. JETP} {\bf 50}(2), 213 (1979).

\bibitem{lk63}
E.M. Lifshitz and I.M. Khalatnikov, 
{\it Adv. Phys.} {\bf 12}(46), 185 (1963).

\bibitem{b87}
J.D. Barrow,
{\it Phys. Lett. B} {\bf 183}(3/4), 285 (1987).

\bibitem{b-arxiv}
B. Li and J.D. Barrow, \emph{Does Bulk Viscosity Create a Viable Unified Dark Matter Model?}, arXiv:0902.3163.

\bibitem{paddy}
T. Padmanabhan and S.M. Chitre, 
{\it Phys. Lett. A} {\bf 120}(9), 433 (1987).

\bibitem{b88}
J.D. Barrow,
{\it Nucl. Phys. B} {\bf 310}(3/4), 743 (1988).

\bibitem{landau-fluid}
L.D. Landau and E.M. Lifshitz,
{\it Course of Theoretical Physics, Volume 6: Fluid Mechanics}
(Pergamon Press, Oxford (UK), 1987).

\bibitem{kolb-turner}
E.W. Kolb and M.S. Turner,
{\it The Early Universe} (Addison-Wesley, Redwood City, California (USA), 1990).

\end{thebibliography}
\end{document}